\documentclass{aa}
\usepackage{graphicx}
\usepackage{txfonts}
\usepackage{hyperref}
\begin{document}

\title{Fluorescent Fe K line emission of $\gamma$~Cas stars}
\subtitle{I. Do $\gamma$~Cas stars host propelling neutron stars?}
\author{G.\ Rauw\inst{1}}
\offprints{G.\ Rauw}
\mail{g.rauw@uliege.be}

\institute{Space sciences, Technologies and Astrophysics Research (STAR) Institute, Universit\'e de Li\`ege, All\'ee du 6 Ao\^ut, 19c, B\^at B5c, 4000 Li\`ege, Belgium}
\date{Received date/Accepted date}

\abstract
{Gamma~Cas stars are early-type Be stars that exhibit an unusually hard and bright thermal X-ray emission. One of the proposed scenarios to explain these properties postulates the existence of a neutron star companion in the propeller stage, during which the magnetosphere of a rapidly rotating neutron star repels infalling material.}
{To test this model, we examined the fluorescent Fe K$\alpha$ emission line at $\sim 6.4$\,keV in the X-ray spectra of $\gamma$~Cas stars, which offers a powerful diagnostic of both the primary source of hard X-rays and the reprocessing material.}
{We computed synthetic line profiles of the fluorescent Fe K$\alpha$ emission line in the framework of the propelling neutron star scenario. Two reservoirs of material contribute to the fluorescence in this case: the Be circumstellar decretion disk and a shell of cool material that surrounds the shell of X-ray-emitting plasma around the putative propelling neutron star.}
{We analysed the synthetic line profiles and expected equivalent widths of the lines for three well-studied $\gamma$~Cas stars. The predicted line strengths fall short of the observed values by at least an order of magnitude. Pushing the model parameters to reproduce the observed line strengths led to column densities towards the primary X-ray source that exceed the observationally determined values by typically a factor of 20, and would further imply a higher X-ray luminosity than observed.}
{The strengths of the observed Fe K$\alpha$ fluorescent emission lines in $\gamma$~Cas stars are inconsistent with the expected properties of a propeller scenario as proposed in the literature.}

\authorrunning{G.\ Rauw}
\titlerunning{$\gamma$~Cas stars}
\maketitle

\section{Introduction}
The unusually hard and bright X-ray emission of the B0.5\,IVe star $\gamma$~Cassiopeiae has been a puzzle ever since its discovery (see \citealt{Smi16} for a review and \citealt{Rau22} for a recent observational study). Over the last two decades, a number of similar objects have been identified \citep[e.g.][]{Lop06,Naz18,Naz20,Naz22,Naz23}, and $\gamma$~Cas is now the prototype of a steadily growing class of over 30 (currently) known objects \citep{Naz23}. All of them are early Be or late Oe stars. They display $\log{L_{\rm X}}$ values (in the 0.5 -- 10\,keV band) ranging from $\sim 31.6$ to $33.2$, $\log{(L_{\rm X}/L_{\rm bol})}$ between $\sim -6.2$ and $-4$, and their X-ray emission is due to a thermal plasma with a temperature $kT \geq 5$\,keV.

Several scenarios have been formulated to explain this X-ray emission. Historically, the first explanation that was proposed was accretion onto a neutron star (NS) companion \citep{Whi82}. However, the absence of pulsations and X-ray bursts as well as the thermal nature and luminosity of the X-ray emission are arguments against this scenario. \citet{Pos17} nevertheless revived the idea of a NS companion. They suggested that the properties of $\gamma$~Cas could be explained provided the companion were a rapidly spinning magnetic NS. This object would then be in the propeller regime, where the rapid rotation of the magnetosphere prohibits accretion. Accretion onto a white dwarf companion was proposed by several authors as an alternative to the NS scenario \citep[e.g.][and references therein]{Mur86,Tsu23,Gie23}. To power the observed level of X-ray emission, this white dwarf scenario would require a rather large mass-transfer rate from the Be star. In view of these difficulties and to account for the observed correlation between X-ray and UV variability, \citet{Smi98} and \citet{Smi99} elaborated another scenario in which small-scale magnetic fields at the Be star surface interact with a magnetic field generated in the Be disk. Finally, \citet{Lan20} proposed that the hard X-ray emission might come from the collision between the wind of a stripped helium subdwarf (sdO) companion and the circumstellar disk of the Be star. In addition to a number of conceptual problems, this scenario was refuted by observing the X-ray properties of a sample of established and suspected Be + sdO systems \citep{Naz22}.

Much like other Be stars, $\gamma$~Cas and a number of its siblings have been found to be binary systems with orbital periods in the range of several weeks to several months \citep[][and references therein]{Naz22a}. The nature of the companions remains uncertain, but the most likely values of the companion masses allow for stripped helium subdwarfs, white dwarfs, or, in some cases, NSs.

To discriminate between the various scenarios, a different road must be followed. In the present work, we focus on the fluorescent Fe K$\alpha$ emission line near 6.4\,keV seen in the X-ray spectra of $\gamma$~Cas stars \citep[e.g.][and references therein]{Gim15}. The strength of these lines offers a diagnostic of the properties of the illuminating hard X-ray source and of the illuminated material. Moreover, in the future, X-ray microcalorimeters on board the {\it X-ray and Imaging Spectroscopy Mission} \citep[XRISM,][]{Ish22} and {\it Athena} \citep{Bar23} will provide high spectral resolution over the 6 - 7\,keV energy band. This will then allow the detailed morphologies of the Fe K$\alpha$ line to be observed, thereby offering an even more powerful diagnostic. As a first application, the present paper concentrates on the \citet{Pos17} scenario, in which the hard X-rays are thought to arise from a shell of hot material around a NS in the propeller regime. Section\,\ref{sect2} describes the ingredients of our calculations of synthetic fluorescent Fe K$\alpha$ line profiles in the propelling NS scenario. The results of the model applied to the specific case of $\gamma$~Cas are presented in Sect.\,\ref{sect3}, whilst Sect. \ref{sect4} outlines the cases of two other $\gamma$~Cas stars. The implications of our results are discussed in Sect.\,\ref{sect5}. 

\section{The Fe K$\alpha$ fluorescent line \label{sect2}}
The illumination of a cloud of relatively cool gas by a bright source of hard X-rays leads to inner-shell photoionization of the atoms and ions in the cool material. In this process, a photon with an energy above the K-shell ionization edge of the corresponding ion is absorbed, thereby creating a vacancy on the K shell. This vacancy is filled by an electron from the L or M shell decaying to the K shell. The excess energy of this electron is shed either via Auger auto-ionization or by the emission of a fluorescent line photon \citep[e.g.][]{Pal03}. When the de-excitation involves an electron from the L shell or the M shell, the fluorescent photons are named respectively K$\alpha$ or K$\beta$. Fluorescent emission of Fe K$\alpha$ photons is of particular interest, because the corresponding line occurs in a relatively clean spectral domain around 6.4\,keV and because Fe ions have relatively high fluorescent yields\footnote{The fluorescent yield is the probability that inner-shell photoionization is followed by fluorescent line emission (either K$\alpha$ or K$\beta$) rather than Auger auto-ionization \citep[e.g.][]{Pal03}.} near 0.3 \citep{Kal04}. In X-ray astrophysics, the Fe K$\alpha$ line bears a considerable diagnostic potential for both the properties of the primary source of hard X-rays and the properties of the illuminated gas. Observationally, fluorescent Fe K$\alpha$ lines have been found in various astrophysical environments, including accretion disks around supermassive black holes in active galactic nuclei \citep[e.g.][]{Tan95,Bren06}, accretion disks around NSs in low-mass X-ray binaries \citep[e.g.][]{Cack08}, accretion flows onto white dwarfs \citep[e.g.][]{Ezu99,Xu16}, flaring or outbursting pre-main-sequence stars \citep[e.g.][]{Skin16,Ham10}, and highly active late-type stars \citep[e.g.][]{Dra08,Cze10}. However, with the clear exception of $\gamma$~Cas stars \citep{Gim15} and possibly some Wolf-Rayet stars \citep{Osk12}, fluorescent Fe K$\alpha$ emission is not seen in the X-ray spectra of massive stars \citep[e.g.][for a review on X-ray emission of massive stars]{Rau22b}.  

\subsection{Modelling fluorescent emission}
In $\gamma$~Cas stars, several media provide reservoirs of material that can potentially contribute to the fluorescent line emission. These include cool plasma in the vicinity of the primary source of hard X-rays, the Be circumstellar disk and the Be photosphere. If the fluorescence arises from hard X-rays illuminating an optically thin slab of cold material, then the ensuing line emission strength is essentially proportional to the column density of the slab. For re-processor column densities higher than $10^{24}$\,cm$^{-2}$, this simple relation breaks down, and absorption and scattering effects become important \citep{Kal95}. For a given column density of the fluorescent material, the Fe K$\alpha$ line luminosity scales mostly with the flux of the illuminating source at the position of the fluorescent material. The equivalent width (EW) of the fluorescent Fe K$\alpha$ line is mostly set by the solid angle occupied by the fluorescent material as seen from the illuminating hard X-ray source. In the present work, where we focus on the propeller scenario, we can safely neglect the contribution from the Be photosphere as the source of hard X-rays is typically located at a distance of several tens of Be stellar radii ($a \sim 35\,R_*$ for $\gamma$~Cas), therefore leading to a very small contribution to the EW. 

We provide here an overview of our model calculations for the fluorescent line profile. We started by considering a cell of material at location $\vec{d}$ from the primary X-ray source and at location $\vec{r}$ from the centre of the Be star. The cell has a hydrogen particle density $n_{\rm H}(\vec{r})$ and its volume is $dV$. To compute the iron K-shell ionization in the cell, we integrated the hard X-ray photon flux per unit energy $f(E,\vec{d})$ attenuated by the optical depth ($\tau(E,\vec{d})$) along the path of the incoming ionizing photon times the K-shell ionization cross-section $\sigma(E)$. The hard X-ray spectrum is represented by an APEC optically thin thermal plasma model \citep{Smi01} of solar abundance \citep{Asp09} extended at energies above 100\,keV by a bremsstrahlung continuum. $\tau(E,\vec{d})$ is obtained by multiplying $\sigma(E)$ by the column density of iron atoms along the line joining the cell to the source of hard X-rays. The K-shell photo-ionization cross-sections are modelled using the parametrizations of \citet{VY} for the Fe ion under consideration. The Fe K$\alpha$ emissivity of a cell of material that is exposed to the hard X-ray source (i.e.\ the primary source as seen from the cell is not occulted by the Be star) is given by

\begin{eqnarray}
  j_{K\alpha}(\vec{r},\vec{d}) & = & E_{K\alpha}\,z_{\rm Fe}\,x_{\rm ion}\,\omega_{\rm ion+1}\,n_{\rm H}(\vec{r}) \nonumber \\
  & & \int_{\rm E_{thres}}^{\infty} \sigma(E)\,f(E,\vec{d})\,\exp{(-\tau(E,\vec{d}))}\,dE  \label{jline}
,\end{eqnarray}where $E_{\rm thres}$ is the threshold energy for K-shell ionization \citep{VY}, $E_{K\alpha}$ is the energy of the K$\alpha$ line \citep{Yam14}, $z_{\rm Fe}$ is the abundance of iron with respect to hydrogen \citep{Asp09}, $x_{\rm ion}$ is the relative abundance of a specific Fe ion, and $\omega_{\rm ion+1}$ the associated K$\alpha$ fluorescent yield \citep{Kal04}. 

The contribution of the cell to the total line luminosity is then evaluated as
\begin{equation}
  dL_{K\alpha}(\vec{r},\vec{d}) = j_{K\alpha}(\vec{r},\vec{d})\,\exp{(-\tau(E_{K\alpha},\vec{r},\vec{d},\vec{n}))}\,dV\,{\rm obs}(\vec{r},\vec{d},\vec{n})
,\end{equation}
where $j_{K\alpha}$ given by equation (\ref{jline}) is expressed in units erg\,s$^{-1}$\,cm$^{-3}$, and the ${\rm obs}(\vec{r},\vec{d},\vec{n})$ parameter is equal to 1 or 0 depending on whether or not the cell can be seen by the observer located in the direction $\vec{n}$. The quantity $\tau(E_{K\alpha},\vec{r},\vec{d},\vec{n})$ yields the optical depth at the energy of the K$\alpha$ line from the position $(\vec{r},\vec{d})$ along the direction $\vec{n}$.

To compute synthetic line profiles as they would be seen by an external observer, we distinguished between the individual line profiles and the morphology of the full Fe K$\alpha$ complex. Indeed, we must keep in mind that what we call the K$\alpha$ line is actually an unresolved array of transitions that arise from the various combinations of L and K shell electronic configurations \citep{Pal03}. Each of these transitions has its own energy and intrinsic strength. The resulting morphology of the Fe K$\alpha$ line is thus obtained by convolving the energy distribution of the transitions with the individual line profile weighted by the intrinsic strength of the transitions. For the computation of the individual line profile, we accounted for the Doppler shift of each cell of fluorescent material and built a histogram of the line emission in radial velocity bins of 50\,km\,s$^{-1}$. Doppler shifts arise for instance from the Keplerian rotation of the material in the circumstellar Be disk. As the illuminating source (in the present scenario associated with the NS) orbits around the Be star, the Doppler shift of those parts of the Be disk that are closest to the source, and thus produce the strongest contribution to the fluorescent line, varies. Therefore, the contribution of the Be disk to the individual line profile is expected to change its morphology as a function of orbital phase. 

\subsection{Fluorescence from the circumstellar Be disk \label{Bedisk}}
The decretion disk of the Be star provides an important reservoir of relatively cool material that can contribute to the formation of the Fe K$\alpha$ line. In the scenario that we test here, the disk is illuminated by a roughly point-like source associated with the hot shell around the NS's magnetosphere. To model the fluorescence from the disk, a number of input parameters are needed. These include the properties of the hard X-ray emission, estimates of the mass and radius of the Be star, the density profile of the circumstellar disk, the disk's inclination to our line of sight, as well as orbital parameters of the binary system.

The X-ray spectrum of $\gamma$~Cas is described well by a combination of up to four optically thin thermal plasma components, with the hottest plasma providing the dominant contribution. For this hot component, \citet{Rau22} determined a mean plasma temperature of $kT = 12.5$\,keV and a mean flux, corrected for absorption by the interstellar medium, of $1.5\,10^{-10}$\,erg\,cm$^{-2}$\,s$^{-1}$ in the 2.0 - 10.0\,keV energy band. To convert the fluxes into luminosities, we adopted a distance of 190\,pc \citep{Naz18}. The fluorescent Fe\,K$\alpha$ line observed in the {\it XMM-Newton} and {\it Chandra} spectra of $\gamma$~Cas has an EW in the range between 34 and 59\,eV \citep{Rau22}.

The mass of the B0.5\,IVe star is usually estimated between 15 and 17\,M$_{\odot}$ \citep{Ste98,Sig07}. We thus adopted a value of 16\,M$_{\odot}$. Estimates of the stellar radius range between 8.2 and 10\,R$_{\odot}$ \citep{Qui97,Ste98,Sig07,Ste12,Kle17}, with a majority of values at 10\,R$_{\odot}$.

There is considerably more dispersion among the literature values of the disk parameters. This could result from either a genuine epoch-dependence of the disk properties or degeneracies of the model parameters. For the disk inclination, the literature values, based either on interferometric observations or fitting of the H$\alpha$ line profile, range between $42^{\circ}$ \citep{Ste12} and $59^{\circ}$ \citep{Sig20}. The majority of the values gather around $45^{\circ}$ \citep{Qui97,Ste98,Sil10} and we thus adopted this value. With the stellar mass and radius considered here, the Keplerian rotation speed at the inner edge of the disk amounts to $v_{\rm rot}(R_*) = 550$\,km\,s$^{-1}$, in agreement with the $(550 \pm 50)$\,km\,s$^{-1}$ obtained by \citet{Ste12} using interferometry. 
For the hydrogen number density distribution of the decretion disk, we adopted the commonly used power-law relation \citep[e.g.][]{Hum00}
\begin{equation}
  n_{\rm H}(r,z) = n_0\,\left(\frac{r}{R_*}\right)^{-\alpha}\,\exp{\left[-\frac{1}{2}\,\left(\frac{z}{H(r)}\right)^2\right]}
,\end{equation}
where $r$ and $z$ are the radial and vertical coordinates of the axisymmetric disk, $n_0$ is the hydrogen number density at the inner edge of the disk, and $H(r)$ is the disk scale height, which varies with $r$ as
\begin{equation}
  H(r) = \frac{c_s(R_*)\,R_*}{v_{\rm rot}(R_*)}\,\left(\frac{r}{R_*}\right)^{\frac{3-p}{2}}
  ,\end{equation}
with $c_s(R_*)$ the speed of sound at the inner disk edge and $p$ {is the power-law exponent of the radial temperature profile of the disk as introduced by \citet{Kur14}. According to these authors, $p$  should be in the range 0 to 0.5. Here we adopt $p = 0.25$.} 
Assuming solar chemical composition, the density determinations in the literature correspond to $n_0$ in the range $3.6\,10^{12}$\,cm$^{-3}$ and $3.6\,10^{14}$\,cm$^{-3}$ \citep{Ste98,Sig07,Sil10,Grz13,Kle17,Sig20}. This rather wide range probably reflects a mix of genuine variability, model degeneracy, and model limitations. Discarding the two most extreme values yields a mean of $2.2\,10^{13}$\,cm$^{-3}$, which we adopted as our default value. For $\alpha$, one can find values ranging from $1.5$ to $3.95$ \citep{Sil10,Grz13,Kle17,Sig20}. We adopted $\alpha = 2.5$ as our default value.
In low-eccentricity Be binary systems, the circumstellar disk is expected to be truncated by gravitational interaction with the companion at the 3:1 resonance radius \citep{Oka01}. In the case of $\gamma$~Cas, this resonance radius corresponds to about 176\,R$_{\odot}$ \citep{Rau22}, which we adopted as the outer radius of the Keplerian disk.   

The binary orbit of $\gamma$~Cas has a period of 203.523\,d and is nearly circular \citep{Nem12,Smi12}. The semi-amplitude of the radial velocity curve of the Be star amounts to 4.1\,km\,s$^{-1}$ and leads to a mass function of 0.00146\,M$_{\odot}$ \citep{Rau22}. Adopting a mass of 16\,M$_{\odot}$ for the Be star and an orbital inclination of $45^{\circ}$ yields a mass estimate of 1.1\,M$_{\odot}$ for the companion and an orbital separation of $a = 375$\,R$_{\odot}$ \citep{Rau22}. This corresponds to a mass ratio $q = \frac{m_{\rm NS}}{m_{\rm Be}} = 0.07$. Using the formalism of \citet{Egg83}, we estimated the radii of the Roche lobe of the binary components to be $0.61\,a$ for the Be star and $0.18\,a$ for the NS. Our default model parameters for $\gamma$~Cas are summarized in Table\,\ref{tab_param}. Throughout this paper, we assume that the companion's orbital plane coincides with the Be star's equatorial plane, which itself contains the circumstellar Be disk. 

\begin{table}
  \caption{Model parameters for three $\gamma$~Cas binary systems. \label{tab_param}}
  \begin{center}
  \begin{tabular}{l c c c}
    \hline
    & $\gamma$~Cas & $\pi$~Aqr & $\zeta$~Tau \\
    \hline
    \multicolumn{4}{c}{Primary X-ray emission}\\
    $kT$ (keV) & 12.5 & 11.5 & 9.0 \\
    $f_{\rm X}$ ($10^{-11}$\,erg\,cm$^{-2}$\,s$^{-1}$) & 15.0 & 1.1 & 1.8 \\
    \multicolumn{4}{c}{Be star}\\
    $M_*$ (M$_{\odot}$) & 16 & 12.7 & 11.2 \\
    $R_*$ (R$_{\odot}$) & 10.0 & 6.5 & 7.7 \\
    $T_{\rm eff}$ (kK) & 25 & 25 & 20 \\
    \multicolumn{4}{c}{Be disk} \\
    $i$ ($^{\circ}$) & 45 & 70 & 85 \\
    $n_0$ ($10^{13}$\,cm$^{-3}$) & 2.2 & 0.45 & 5.8 \\
    $\alpha$ & 2.5 & 2.5 & 3.0 \\
    $R_{\rm out}$ (R$_{\odot}$) & 176 & 91 & 118 \\
    \multicolumn{4}{c}{Binary orbit} \\
    $P_{\rm orb}$ (d) & 203.523 & 84.1 & 132.987 \\
    $a$ (R$_{\odot}$) & 375 & 193 -- 198 & 252 \\
    \hline 
  \end{tabular}
  \end{center}
\end{table}

\begin{figure}
  \resizebox{9cm}{!}{\includegraphics[angle=0]{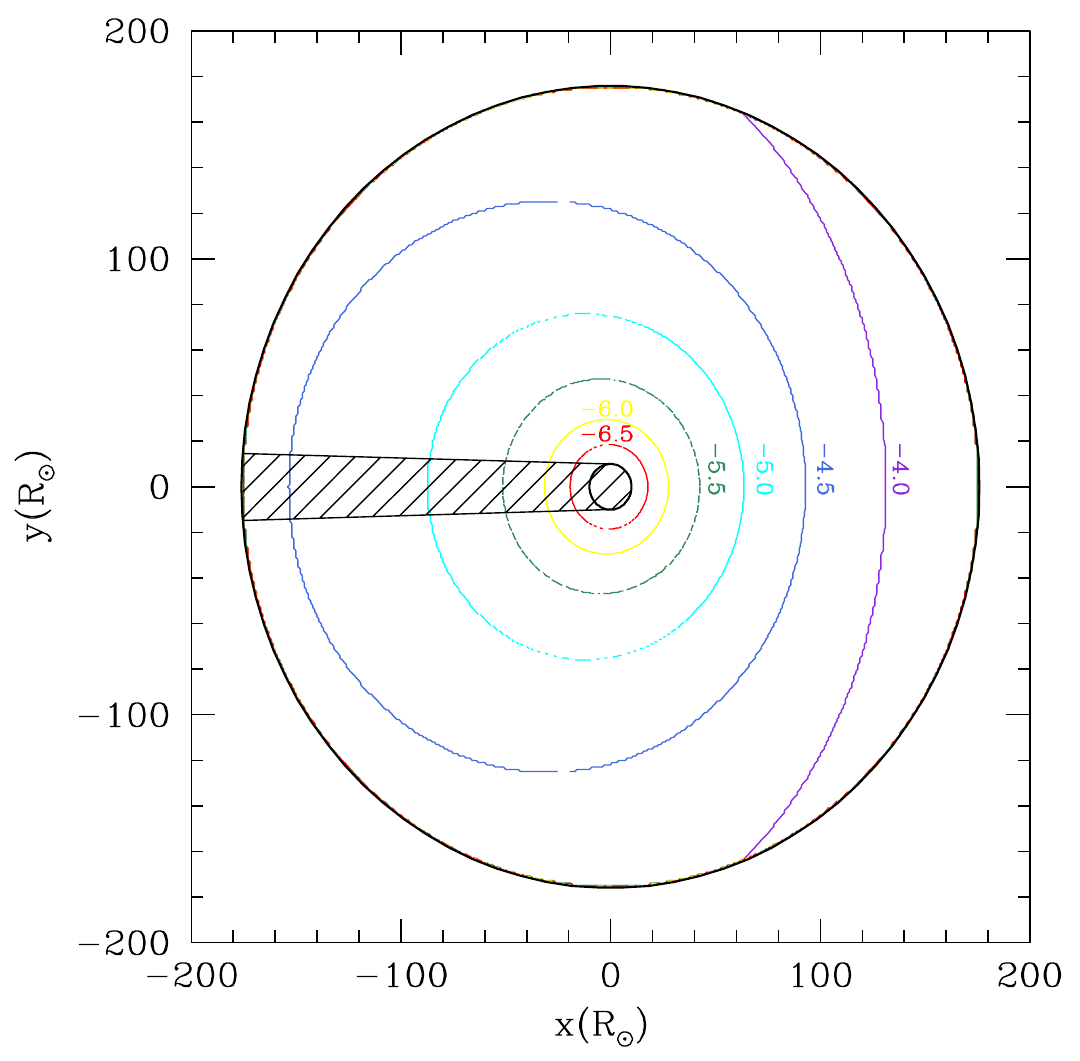}}
  \caption{Distribution of the ionization parameter, $\xi$, over the Be decretion disk for our $\gamma$~Cas default model (Table\,\ref{tab_param}). The hard X-ray source is located to the right at $x = 375$\,R$_{\odot}$, $y=0$. The labels on the different contours correspond to the value of $\log{\xi}$ (from $-4.0$ for the blue-violet contour to $-6.5$ for the red contour). The hatched area lies in the shadow of the Be star and is thus not illuminated by the hard X-ray source. \label{xi}}
\end{figure}

The presence of a hard X-ray source near the disk could in principle alter the ionization of the disk material. This is expressed via the $\xi$ parameter defined as
\begin{equation}
  \xi = \frac{4\,\pi\,F_{\rm X}}{n_e}
  \label{xiEquation}
,\end{equation}
where $n_e$ is the electron density and $F_{\rm X}$ is the X-ray flux over the 1 - 1000\,Ry energy domain \citep[e.g.][]{Kal04}. Figure\,\ref{xi} illustrates the distribution of this ionization parameter over the Be decretion disk. The value of $\xi$ inside the disk is the result of two competing effects. On the one hand, the $r^{-\alpha}$ dependence of $n_e$ with distance from the centre of the Be star implies higher $\xi$ values as $r$ increases for a given $F_{\rm X}$. On the other hand, the dependence of $F_{\rm X}$ on $d^{-2}$ where $d$ is the distance from the source of hard X-rays (located here at  $x = 375$\,R$_{\odot}$, $y=0$), leads to a reduction of $\xi$ as we move away from the NS's position. The largest values of $\xi$ are thus logically expected in the outer areas of the Be disk facing directly the source of hard X-rays. Our calculations show that the highest values of $\log{\xi}$ is expected to be about $-4.0$. Given this low value of $\xi$, the ionization state of the Be disk is not strongly impacted by the presence of the hard X-ray source \citep{Kal04}. 

The vertical column density of the Be disk, measured from the disk plane where most of the fluorescence takes place, is given by 
\begin{eqnarray}
  \int_{0}^{+\infty} n_{\rm H}(r,z)\,dz & = & n_0\,\left(\frac{r}{R_*}\right)^{-\alpha}\,H(r)\,\sqrt{\frac{\pi}{2}} \nonumber \\
  & = &n_0\,\left(\frac{r}{R_*}\right)^{\frac{3-p}{2}-\alpha}\,\frac{c_s(R_*)\,R_*}{v_{\rm rot}(R_*)}\,\sqrt{\frac{\pi}{2}}
.\end{eqnarray}
Because $\frac{3-p}{2}-\alpha$ is negative for all values of $\alpha$ that have been proposed for $\gamma$~Cas, the maximum value of the column density occurs at the inner edge of the disk and amounts to $n_0\,\frac{c_s(R_*)\,R_*}{v_{\rm rot}(R_*)}\sqrt{\pi/2}$. With the default model parameters of Table\,\ref{tab_param}, this quantity amounts to $6.3\,10^{23}$\,cm$^{-2}$, making the inner disk mildly optically thick. In the present case, most of the fluorescence occurs however in the outer disk regions where the column density is significantly lower and the relevant parts of the disk are thus optically thin.

\subsection{Fluorescence from the cool propeller shell}
\citet{Pos17} proposed that the $\gamma$~Cas phenomenon could be explained if the companion of the Be star is a rapidly rotating magnetic NS in the so-called propeller stage. In this situation, direct accretion of material onto the NS is inhibited by the centrifugal barrier of the NS's rigidly rotating magnetosphere, which at its outer boundary rotates faster than the Keplerian velocity. As a result, the gravitationally captured material from the Be star outflow would accumulate in a shell around the magnetosphere. \citet{Pos17} infer a gravitational luminosity of the shell equal to
$L_{\rm X} = \frac{G\,M_{\rm NS}\,\dot{M}_{\rm BHL}}{R_{\rm A}}$
where $M_{\rm NS}$ is the mass of the NS, $\dot{M}_{\rm BHL}$ is the rate at which mass is accumulated via the Bondi-Hoyle-Lyttleton mechanism, and $R_{\rm A}$ is the NS's Alfv\'en radius. At $R_{\rm A}$, for which \citet{Pos17} propose a value of $10^4$\,km, the shell material would reach temperatures of order $T_{\rm A} \simeq 100$\,MK as the infalling material bounces against the rigidly rotating magnetosphere.

This shell of gravitationally captured material should extend between $R_{\rm A}$ and the Bondi-Hoyle-Lyttleton radius for capture of material from the Be outflow $R_{\rm BHL} = \frac{2\,G\,M_{\rm NS}}{v_0^2}$. Here $v_0$ is the relative velocity at which the material from the Be wind or disk outflow encounters the NS. For a NS mass near 1.1\,M$_{\odot}$ and considering $v_0 \sim 100$\,km\,s$^{-1}$ \citep{Smi17}, $R_{\rm BHL}$ amounts to about $\sim 2.9\,10^7$\,km or 42\,R$_{\odot}$.

\citet{Pos17} consider that the material in the shell behaves as an adiabatic ideal gas. This assumption then leads to a radial profile of the temperature and density that decreases outwards as
\begin{equation}
  T(s) = T_{\rm A}\,\frac{R_{\rm A}}{s}
  \label{Tempstrat}
\end{equation}
and
\begin{equation}
  n_e(s) = n_{e, \rm A}\,\left(\frac{R_{\rm A}}{s}\right)^{2/3}
,\end{equation}
where $s$ is the radial distance from the NS.

\begin{figure}
  \resizebox{8cm}{!}{\includegraphics[angle=0]{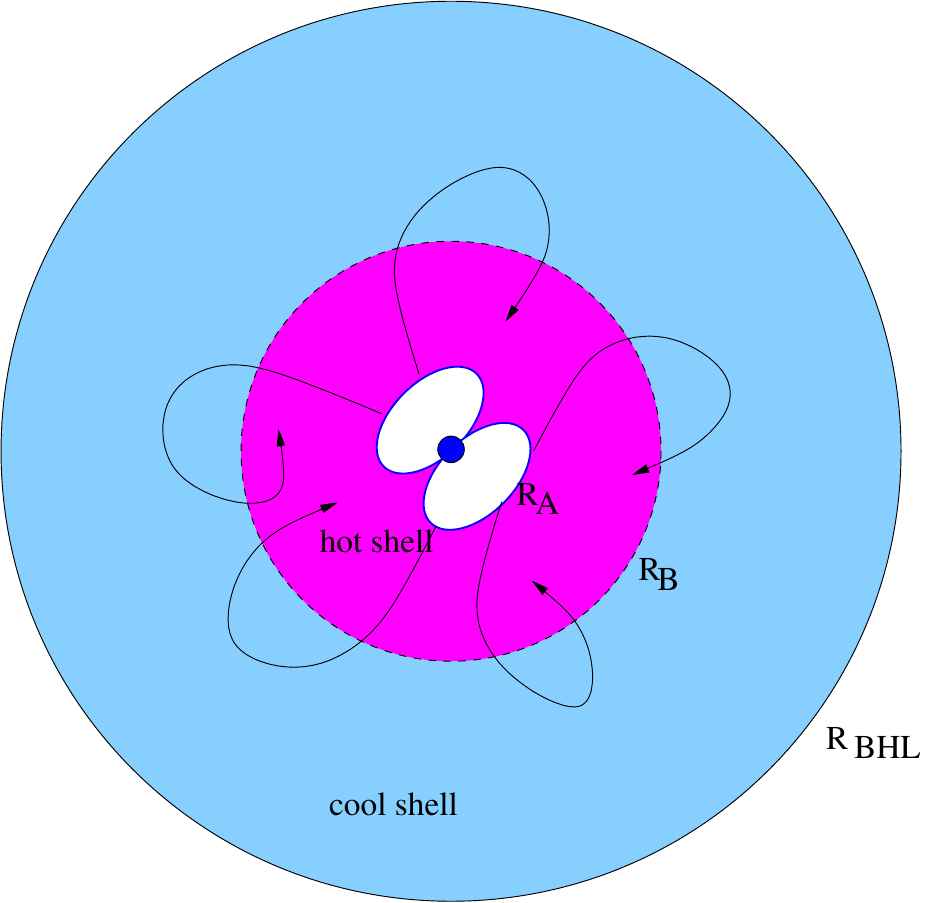}}
  \caption{Schematic illustration of the shells of hot and cool material in the propeller NS scenario. The different radii ($R_{\rm A}$, $R_{\rm B}$ and $R_{\rm BHL}$) are not to scale. The curved trajectories illustrate the motion of heated particles as they drift outwards over the cooling time. \label{propeller}}
\end{figure}

\citet{Pos17} further consider that the gas in the shell moves in turbulent motion at a velocity
\begin{equation}
  v_{\rm turb}(s) = \epsilon\,\sqrt{\left(\frac{\gamma\,k\,T(s)}{m}\right)}
  \label{turb}
,\end{equation}
with $\epsilon \leq 1$, $\gamma = \frac{5}{3}$, and $m$ the mean mass of particle. \citet{Pos17} adopt $v_{\rm turb} = 1000$\,km\,s$^{-1}$ for the hottest gas\footnote{\citet{Smi17} pointed out that this value of $v_{\rm turb}$ is a factor of 2 -- 3 higher than the actual line broadening measured in high-resolution X-ray spectra of $\gamma$~Cas, suggesting that the value of $v_{\rm turb}$ might have to be reduced to below $1000$\,km\,s$^{-1}$.}. Over the radiative cooling time of $\sim 1000$\,s, the hot gas would therefore travel out to a distance of order $R_{\rm B} \sim 100\,R_{\rm A} \sim 10^6$\,km. This distance is significantly (29 times) smaller than $R_{\rm BHL}$. We can thus assume that the hot plasma, which would constitute the primary X-ray source located between $R_{\rm A}$ and $R_{\rm B}$, should be surrounded by a second shell of cooler gas ($T < 1$\,MK) possibly extending out to $R_{\rm BHL}$. A schematic view of this situation is provided in Fig.\,\ref{propeller}. {The boundary between these two shells is somewhat arbitrarily set to the radial coordinate $s$ where a temperature of 1\,MK is reached according to Eq.\,\ref{Tempstrat}. In reality, due to the turbulence (see Eq.\,\ref{turb}), several gas temperatures likely coexist over a given range of the radial coordinate $s$.}

{Adopting the value $n_{\rm H,A} = 10^{13}$\,cm$^{-3}$ advocated by \citet{Pos17}, the ionization parameter $\xi$ (Eq.\,\ref{xi}) amounts to 0.22 at the inner boundary of the cool shell ($R_{\rm B}$). This parameter decreases outwards as $s^{-4/3}$, reaching 0.0024 at $R_{\rm BHL}$. According to Fig.\,5 of \citet{Kal04}, photoionization of iron would then result in Fe\,{\sc vi} and Fe\,{\sc vii} being the dominant ionization stages near the inner boundary of the cool shell. Farther out in the cool shell, photoionization would have a significantly lower impact, leaving Fe\,{\sc iv} and Fe\,{\sc v} as the dominant ionization stages. If we instead adopted $n_{\rm H,A} = 5\,10^{14}$\,cm$^{-3}$ (see Sect.\,\ref{shellresult}), then $\log{\xi}$ would be reduced by $-1.70$ and photoionization plays no significant role in setting the ionization stage of iron. Whilst we assume here for simplicity that the fluorescent emission involves ions of a single ionization stage, this assumption has no impact on our main conclusions regarding the overall strength of the fluorescent line.}

We thus considered the contribution of this putative cool shell to the fluorescent emission. For this purpose, we discretized the cool shell into 100 sub-shells of linearly increasing radius. {Each of these sub-shells is discretized according to polar angle (in steps of $1^{\circ}$) as measured from the direction of the line of sight towards the observer. Whilst the properties of the cool shell solely depend on the radial coordinate $s$ and are independent of polar angle, the column density along the pathway from each sub-shell element to the observer does depend on polar angle. This column density is integrated numerically to compute the optical depth $\tau$ at the energy of the Fe K$\alpha$ photons.} The total fluorescent line emission results from the sum of the emission of each sub-shell element attenuated by the associated $\exp(-\tau)$.  

We can estimate the column density towards the primary X-ray source that this shell of cool material would produce
\begin{eqnarray}
 N_{\rm H} = \int_{R_{\rm B}}^{R_{\rm BHL}} n_{\rm H}\,ds & = &\int_{R_{\rm B}}^{R_{\rm BHL}} n_{\rm H,A}\,\left(\frac{R_{\rm A}}{s}\right)^{3/2}\,ds \nonumber \\
  & = & 2\,n_{\rm H,A}\,R_{\rm A}\,\left(\frac{R_{\rm A}}{R_{\rm B}}\right)^{1/2}\,\left[1 - \sqrt{\frac{R_{\rm B}}{R_{\rm BHL}}}\right] \label{columndens}
.\end{eqnarray}
With $n_{\rm H,A} \simeq 10^{13}$\,cm$^{-3}$ and $R_{\rm A} = 10^4$\,km \citep{Pos17}, we obtain a column density of around $1.6\,10^{21}$\,cm$^{-2}$. This value is comparable to the column densities found empirically in the spectral fitting of the X-ray spectra of $\gamma$~Cas, although the observationally determined column density varies strongly from one epoch to the other \citep{Rau22}. The existence of a shell of cool material around the primary X-ray source is thus not inconsistent with the observations.

\begin{figure*}
  \begin{minipage}{8.5cm}
    \resizebox{8.5cm}{!}{\includegraphics[angle=0]{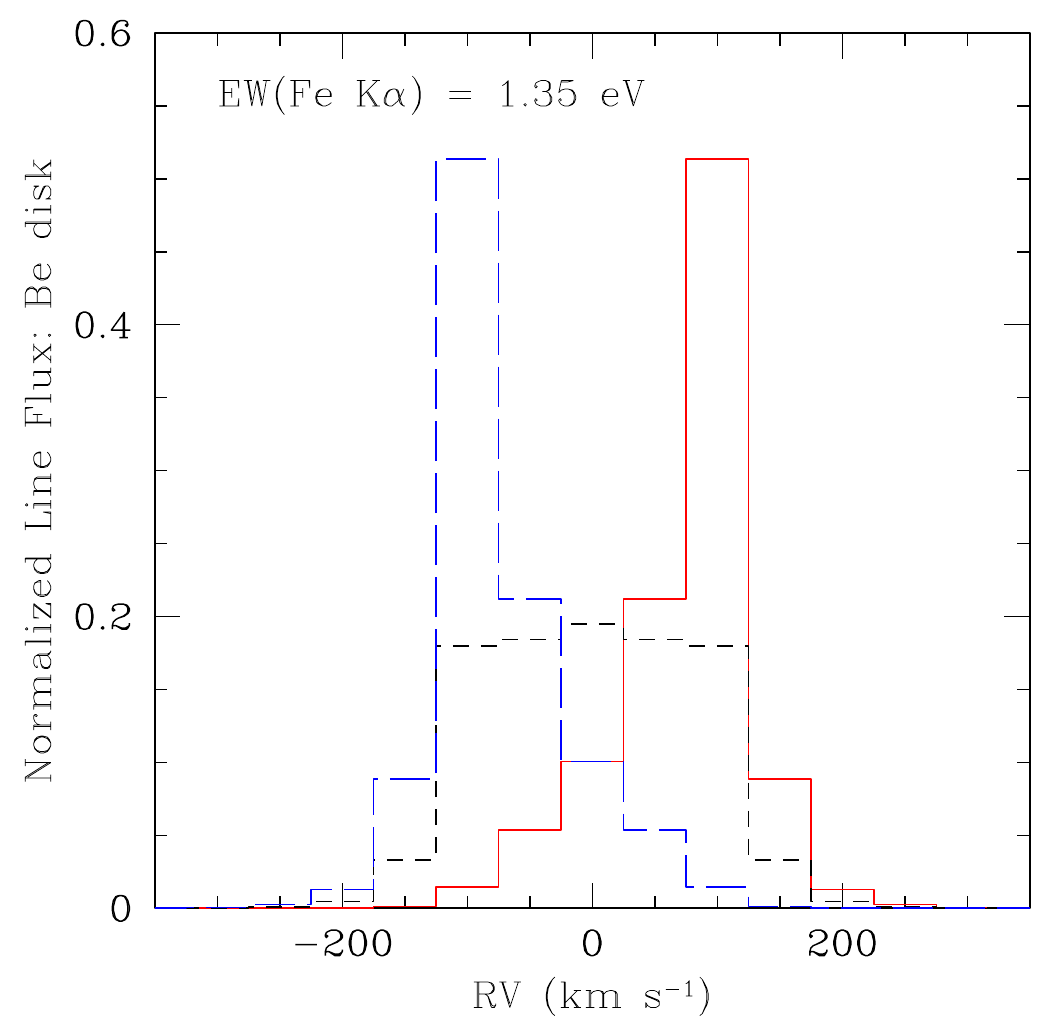}}
  \end{minipage}
  \hfill
  \begin{minipage}{8.5cm}
    \resizebox{8.5cm}{!}{\includegraphics[angle=0]{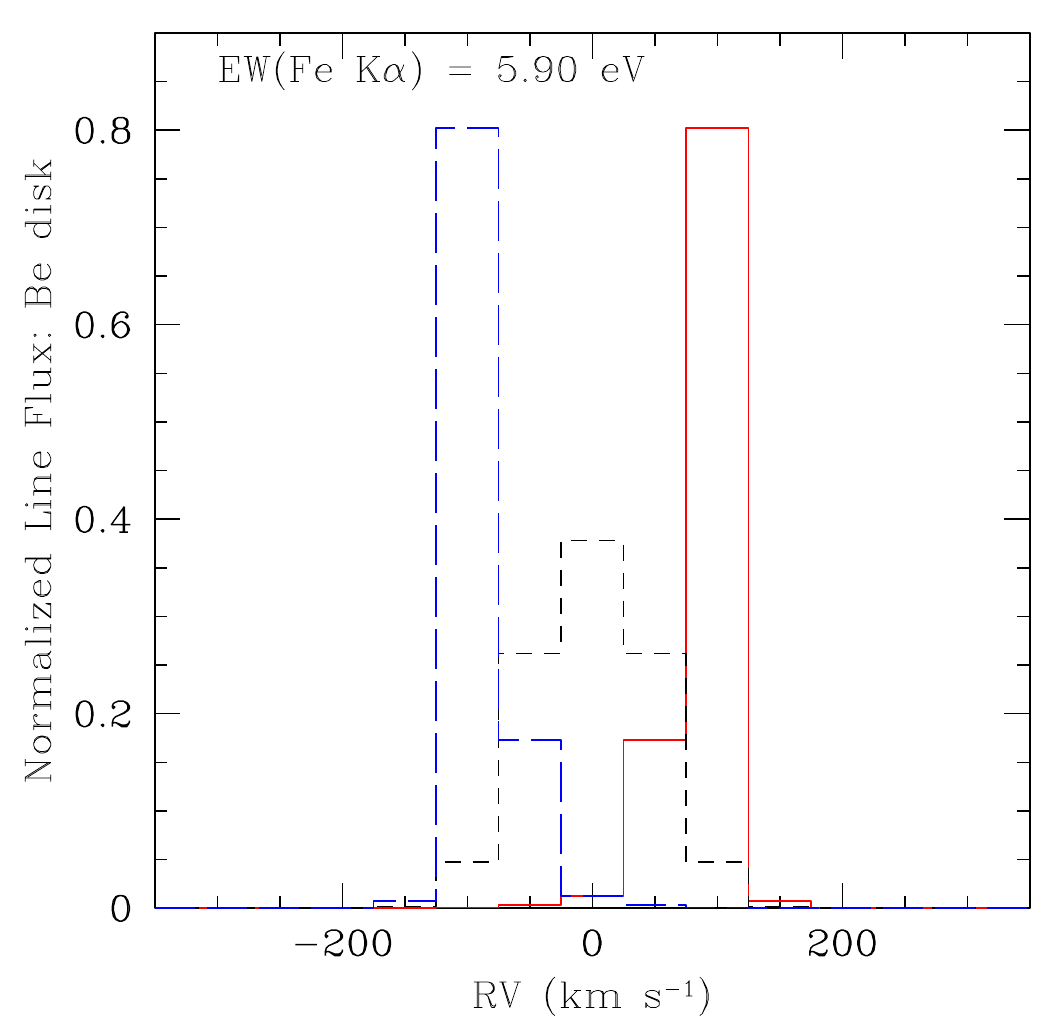}}
  \end{minipage}  
  \caption{Predicted normalized individual Fe K$\alpha$ line profiles for the Be disk. The line profiles are binned into velocity bins of 50\,km\,s$^{-1}$ and are normalized by dividing by the total flux over the entire line profile. The blue, black and red profiles correspond to orbital phases 0.75, 0.00 and 0.25 where phase 0.0 corresponds to the inferior conjunction of the NS. The left and right panels respectively illustrate the profiles obtained with our default set of parameters (Table\,\ref{tab_param}) and with the disk density distribution of \citet{Sil10}. \label{Disk}}
\end{figure*}

As indicated in Eq.\,\ref{turb}, the motion of the gas in the Bondi-Hoyle-Lyttleton volume is assumed to be dominated by turbulence with a turbulent velocity that is a fraction of the local speed of sound. As the temperature decreases outwards, the turbulent velocity should thus decrease from about 100\,km\,s$^{-1}$ at $R_{\rm B}$ to about 20\,km\,s$^{-1}$ at $R_{\rm BHL}$. Hence, we expect the contribution of the cool shell to the Fe K$\alpha$ line profile to a have a relatively narrow shape, consisting of the combination of Gaussian line profiles of decreasing widths as we consider the contribution of layers of larger radii. The centre of this profile is expected to be shifted in velocity by the NS's orbital motion. From the most recent orbital solution of $\gamma$~Cas, we can estimate an amplitude of the orbital motion of the companion of $K_{\rm NS} \simeq 60$\,km\,s$^{-1}$. In our model, we thus assume that the line flux from a layer of radius $s$ in the cool shell is distributed in a Gaussian profile of variance $\sigma^2 = \frac{v_{\rm turb}(s)}{2\,\ln{2}}$ and centred at $K_{\rm NS}\,\sin{(2\,\pi\,\phi)}$. Here $\phi$ is the orbital phase with $\phi = 0$ at the inferior conjunction of the NS star\footnote{\citet{Nem12} defined phase 0.0 as the time of minimum radial velocity of the Be star. Hence, our phases are shifted by 0.25 with respect to theirs: $\phi = \phi_{\rm Nemravova} + 0.25$.}. The contribution of the entire cool shell to the individual line profile is then obtained by combining the contributions from each sub-shell of radius $s$ between $R_{\rm B}$ and $R_{\rm BHL}$. 

\section{Results \label{sect3}}
\subsection{Fe K$\alpha$ emission from the Be disk \label{Bediskresult}}
We discretized the Be disk into 18000 cells, each of them spanning $2^{\circ}$ in azimuthal angle, and extending over a radial distance $dr = (R_{\rm out} - R_*)/100$. Our code computes the contribution of each cell according to equation\,\ref{jline}, accounting for the cell's visibility and illumination by the primary X-ray source. We ran our code assuming seed Fe ions of Fe\,{\sc i}, Fe\,{\sc ii}, Fe\,{\sc iii}, or Fe\,{\sc iv}. The resulting line luminosities and EWs are very similar for each of these ions. In the following we focus on the line profiles and line characteristics obtained assuming Fe\,{\sc iv} seed ions. With our default parameters, we obtain a total line luminosity\footnote{The line luminosities and EWs quoted in this paper refer to the integrated quantities of the entire K$\alpha$ line complex, rather than the values of a specific transition of the array.} of $9.1\,10^{28}$\,erg\,s$^{-1}$ and an EW of 1.35\,eV. The individual line profiles for three different orbital phases are shown in Fig.\,\ref{Disk}. As expected, the line emission from the Be disk displays phase-dependent shifts of the line centroid (by about 200\,km\,s$^{-1}$ peak-to-peak) as well as phase-dependent changes of the full width at half maximum (FWHM). The most important conclusion though is the fact that, with our default parameters, the predicted line EW is much lower than the observed range of EWs \citep[from 34 to 59 eV,][]{Rau22}. {Whilst we focus mostly on the comparison between the observed and synthetic EWs, exactly the same conclusions are reached by comparing observed and synthetic line fluxes. Indeed, in our calculations the continuum radiation against which the EW is computed is evaluated from the model that best fits the observed spectra, thus making both quantities, line flux and EW, equivalent.} 

\begin{figure*}
  \begin{minipage}{8.5cm}
    \resizebox{8.5cm}{!}{\includegraphics[angle=0]{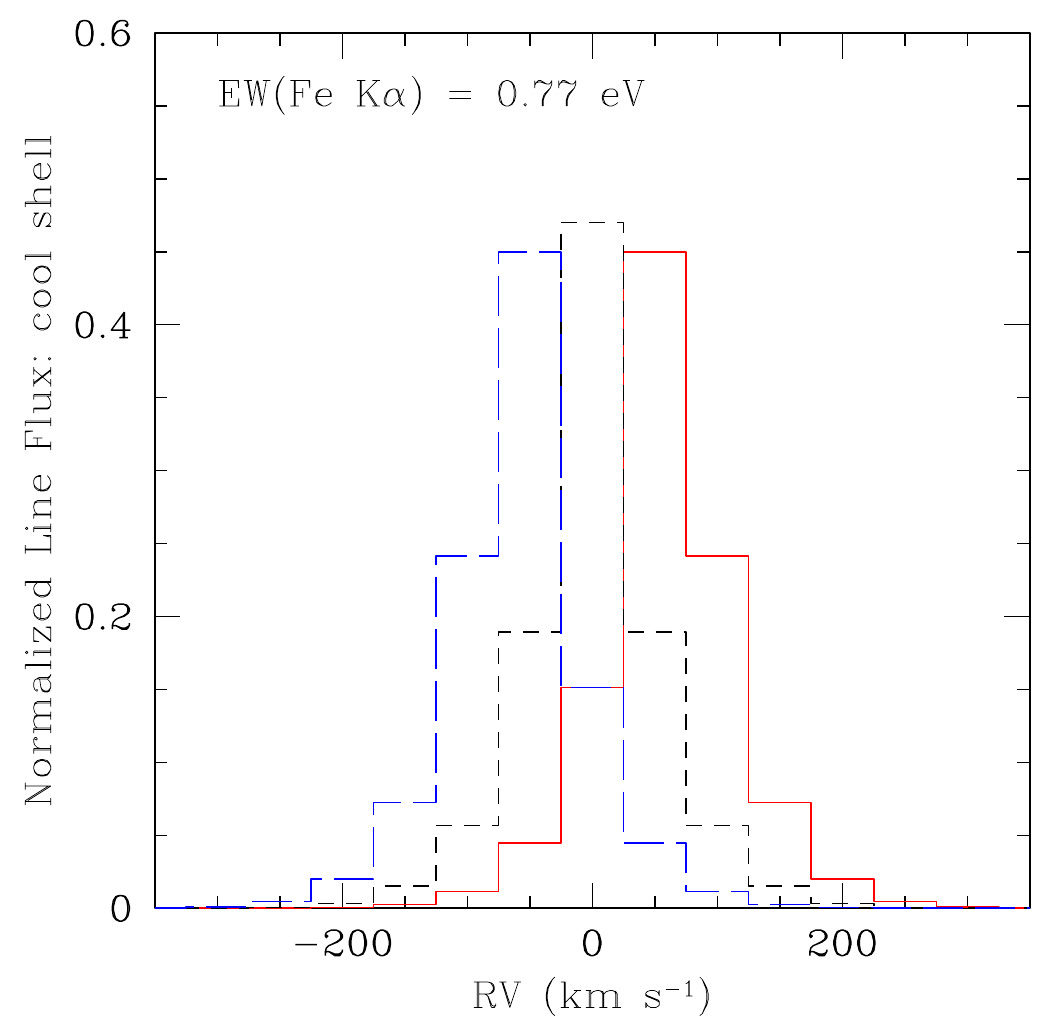}}
  \end{minipage}
  \hfill
  \begin{minipage}{8.5cm}
    \resizebox{8.5cm}{!}{\includegraphics[angle=0]{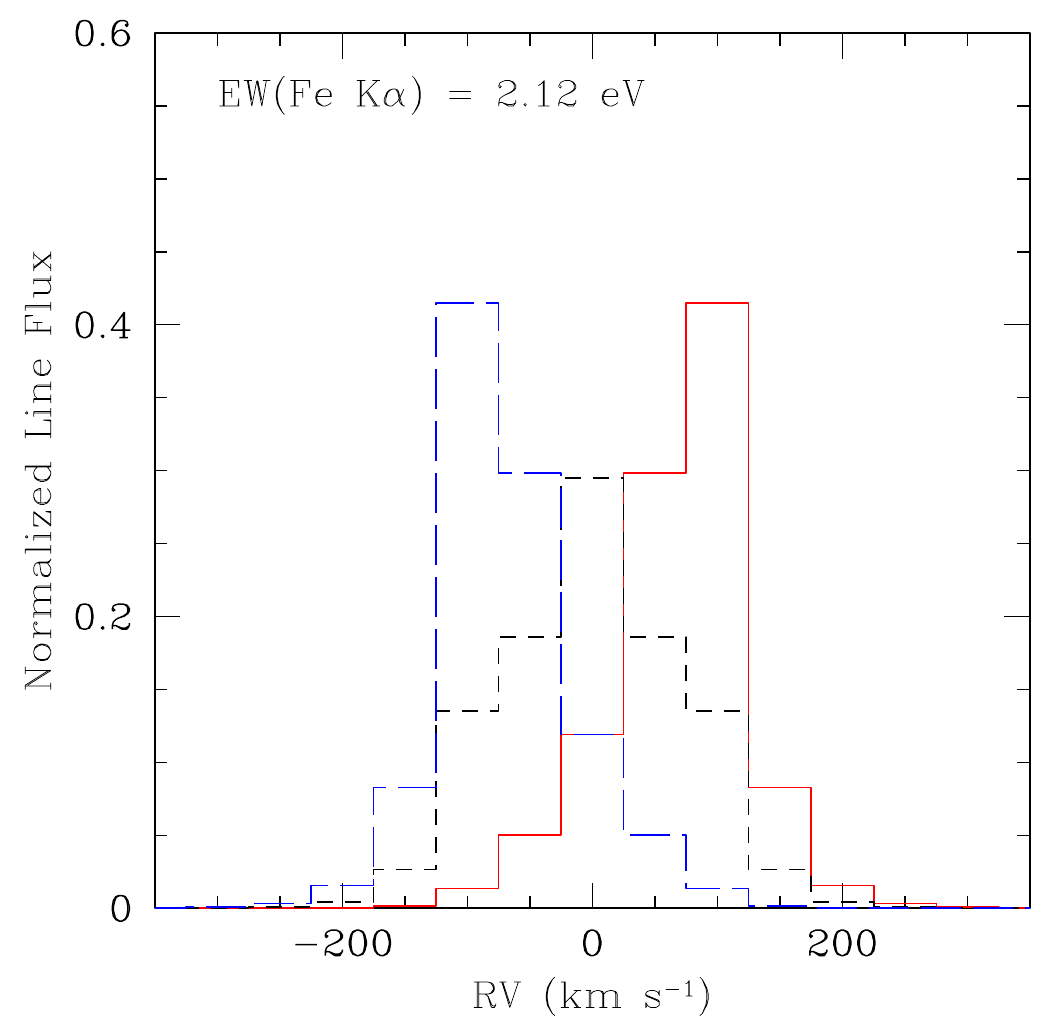}}
  \end{minipage}  
  \caption{Predicted normalized individual Fe K$\alpha$ line profiles for the propeller NS scenario with $n_{\rm  H,A} = 10^{13}$\,cm$^{-3}$. The line profiles are binned into velocity bins of 50\,km\,s$^{-1}$ and are normalized by dividing by the total flux over the entire line profile. The blue, black, and red profiles correspond to orbital phases 0.75, 0.00, and 0.25. The left and right panels respectively display the contribution of the cool shell around the propelling NS and the sum of this contribution and of the emission arising from the Be disk. \label{Postn1}}
\end{figure*}

Given our above discussion about the uncertainties affecting the density profile of the Be disk, we can ask ourselves what would be the impact of a different assumption on $n_0$ and $\alpha$. The highest value of $n_0$ ($3.6\,10^{14}$\,cm$^{-3}$) was proposed by \citet{Grz13}, along with a value of $\alpha = 3.95$ (i.e.\ the steepest density profile proposed for the disk of $\gamma$~Cas). We thus tested this combination of parameters in our model, keeping all other parameters identical. This results in an EW of 0.83\,eV, lower than what we have found with our default parameters. The reason is the high value of $\alpha$, which leads to low densities in the outer disk regions, which are closest to the hard X-ray source. Hence, instead of looking for higher values of $n_0$, we tried a flatter power law. We thus adopted the density distribution proposed by \citet{Sil10} who derived $\alpha = 1.5$ and $n_0 = 4.5\,10^{13}$\,cm$^{-3}$. These parameters lead to an EW of 5.90\,eV, about 4.5 times higher than the value for our default model. The corresponding line profiles are shown in the right panel of Fig.\,\ref{Disk}. Yet, this value of the EW is still well below the observed values. As a last step, we thus assumed that, instead of being truncated at the 3:1 resonance radius, the disk extended out to the radius of the Be star's Roche lobe, namely $R_{\rm out} = 229$\,R$_{\odot}$. For our default density profile, this assumption leads to EW = 2.18\,eV, while we obtain 13.6\,eV when adopting the most extreme density profile \citep{Sil10}. The 13.6\,eV value likely represents an upper limit to the contribution that we can expect from the Be disk in the case of $\gamma$~Cas and assuming the NS propeller scenario.

Whilst \citet{Pos17} attributed the observed Fe K$\alpha$ line to fluorescence from the Be disk, our results show that the EWs of the line contributed by the Be disk would be significantly smaller than the observed values. Therefore, we conclude that, additional contributions to the line are needed if the properties of $\gamma$~Cas are to be explained by a propeller NS scenario. 

\subsection{Fe K$\alpha$ emission from the cool propeller shell \label{shellresult}}
\begin{figure}
  \resizebox{8cm}{!}{\includegraphics[angle=0]{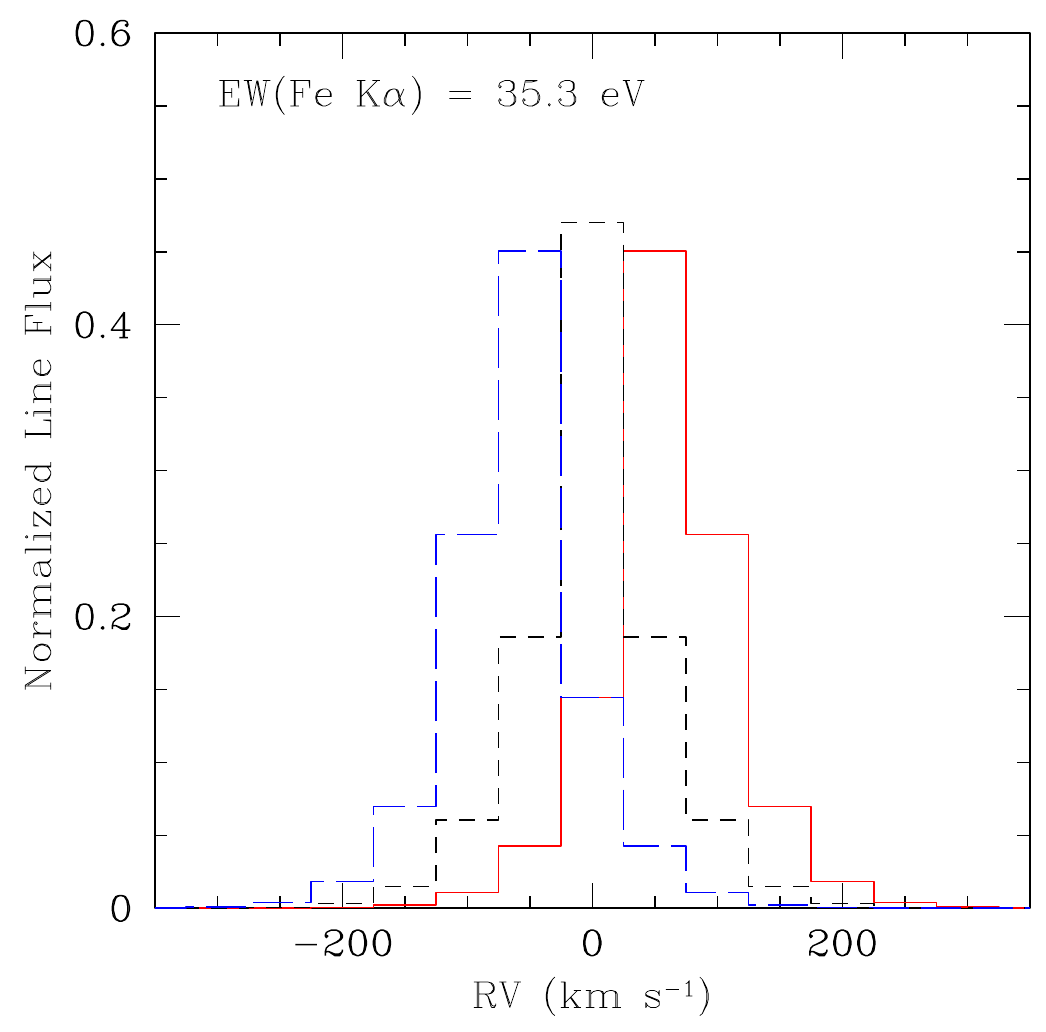}}
  \caption{Same as the right panel of Fig.\,\ref{Postn1}, but for $n_{\rm H,A} = 5.0\,10^{14}$\,cm$^{-3}$. \label{Postn2}}
\end{figure}

As a next step, we considered the contribution to the Fe K$\alpha$ line from the cool shell around the NS's magnetosphere. If we use the parameter $n_{\rm H,A} = 10^{13}$\,cm$^{-3}$ advocated by \citet{Pos17}, we obtain an EW of the shell contribution of 0.77\,eV. Together with the disk contribution for the default parameters this yields a total EW of 2.12\,eV. This is again much lower than the observed line strength. If we rather set $n_{\rm H,A} = 5.0\,10^{14}$\,cm$^{-3}$, the shell contributes 33.9\,eV and the total EW becomes 35.3\,eV. The corresponding individual line profiles are illustrated in Figs.\,\ref{Postn1} and \ref{Postn2}, respectively, for $n_{\rm H,A} = 10^{13}$\,cm$^{-3}$ and $n_{\rm H,A} = 5.0\,10^{14}$\,cm$^{-3}$. 

\begin{figure}
  \resizebox{8cm}{!}{\includegraphics[angle=0]{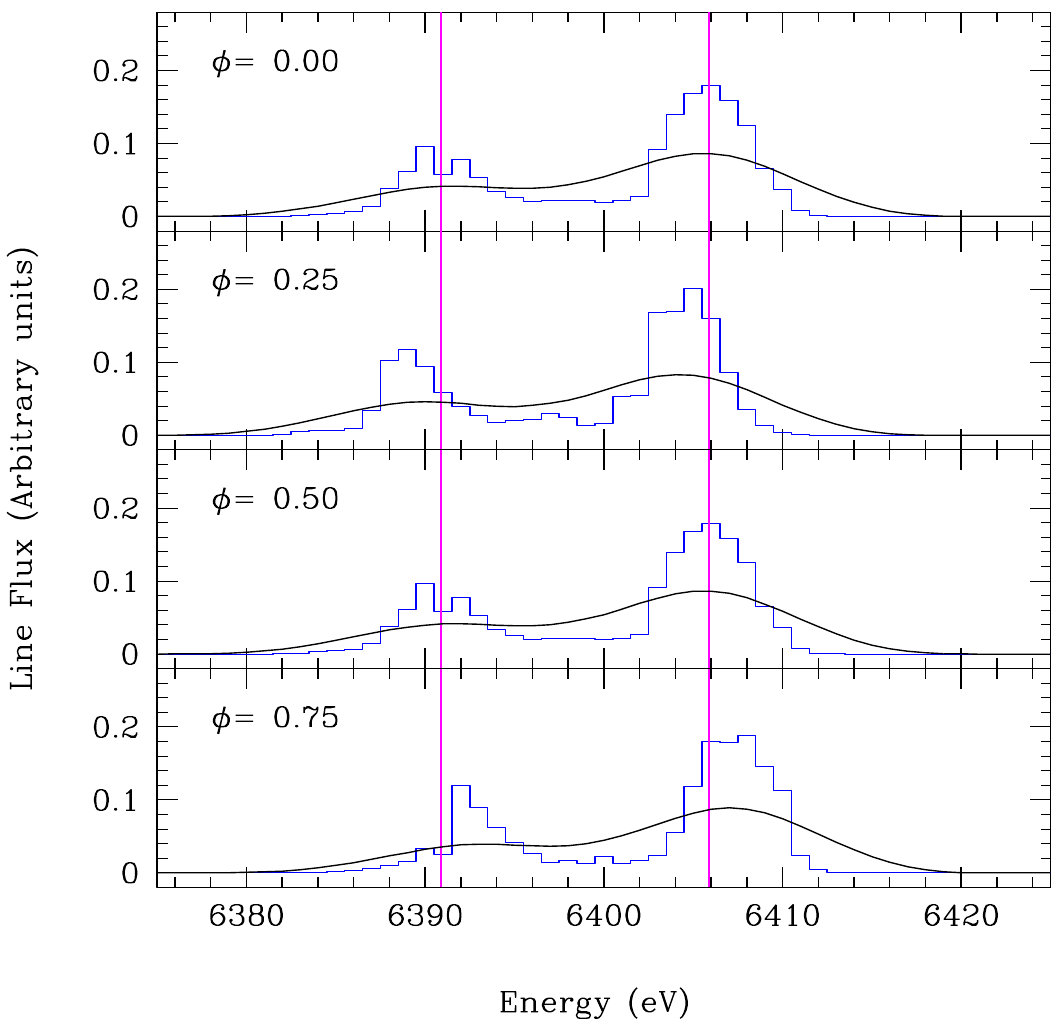}}
  \caption{Illustration of the global line profile (i.e.\ combining the disk and shell contributions and accounting for the full array of transitions), adopting the default parameters of $\gamma$~Cas at four different orbital phases. The blue histogram shows the spectrum binned in energy bins of 1\,eV, whereas the black line provides the result of the convolution with an instrumental response of 6\,eV. The magenta lines indicate the position of the centroids of the K$\alpha_1$ and K$\alpha_2$ subgroups at phase $\phi = 0.0$.\label{global}}
\end{figure}

As pointed out above, the Fe K$\alpha$ feature is not a single line, but instead a complex array of transitions. To illustrate this, the total simulated profiles computed for the set of default parameters are displayed for four orbital phases in Fig.\,\ref{global}. {For the purpose of this figure, we have adopted an Fe\,{\sc iv} ionization stage for the fluorescent material, though we stress that very similar results are obtained for other ionization stages.} In this plot, the blue histograms illustrate the synthetic spectrum with energy bins of 1\,eV. The black curves illustrate the result of a convolution of this synthetic spectrum with a Gaussian whose FWHM is equal to 6\,eV, which roughly mimics the instrumental response of the Resolve X-ray microcalorimeter aboard the XRISM satellite. Whilst the strength of the line simulated here (EW = 2.12\,eV) is well below the actually observed value, it is worth stressing that the expected global shifts of the centroids with orbital phase are very similar if we assume higher values of $n_{\rm H,A}$ or a flatter disk density profile. The amplitude of these shifts are well within the reach of future high-resolution and high-sensitivity instruments such as Resolve on XRISM or the X-ray Integral Field Unit (X-IFU) on {\it Athena}, which should have energy resolutions of 5--7\,eV and $\leq 4$\,eV, respectively.
   
\begin{figure}
  \resizebox{9cm}{!}{\includegraphics[angle=0]{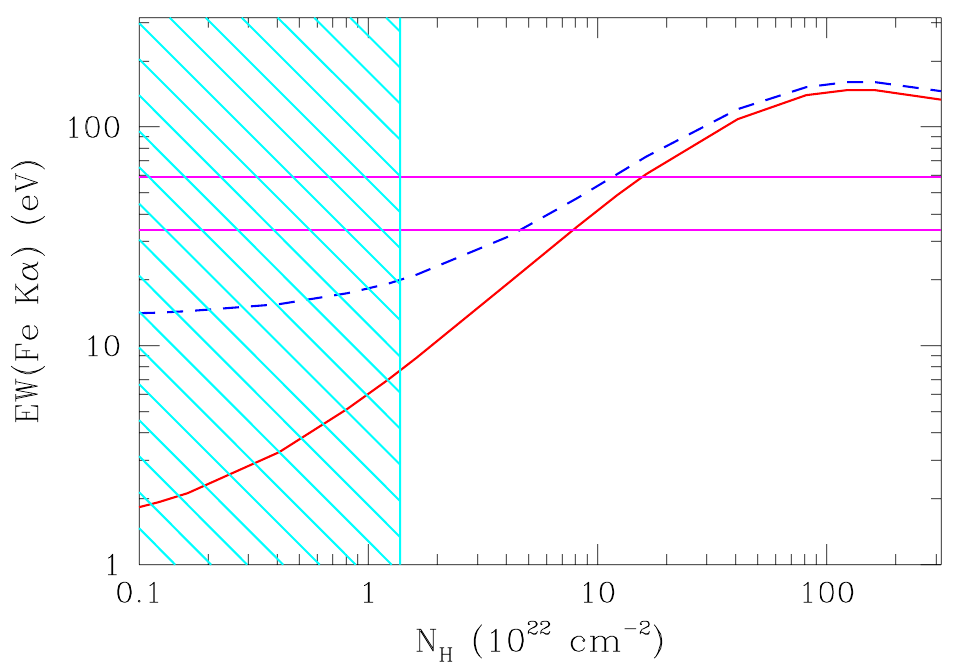}}
  \caption{{Dependence of the total EW(Fe K$\alpha$) from both the cool shell around the propelling NS and the Be disk on the hydrogen column density towards the hot X-ray plasma (evaluated according to Eq.\,\ref{columndens}). The red and dashed blue curves illustrate the total EW(Fe K$\alpha$) computed respectively for the default model parameters and assuming a Be disk with a \citet{Sil10} density profile and extending out to the radius of the Roche lobe. The horizontal magenta lines illustrate the range of the observed EW values. The hatched cyan area illustrates the range of observed hydrogen column densities.} \label{coldens}}
\end{figure}

As we have shown before, the hydrogen column density towards the primary X-ray source scales linearly with $n_{\rm H,A}$ (Eq.\,\ref{columndens}). As long as the cool shell remains optically thin, the EW of the fluorescent line also scales linearly with $n_{\rm H,A}$. {At high densities, the relation between EW and $N_{\rm H}$ deviates however from a straight line because of the increasing optical depths. These relations are illustrated in Fig.\,\ref{coldens}. The horizontal axis of this figure indicates $N_{\rm H}$ in logarithmic scale. The vertical axis yields the EW of the full fluorescent Fe K$\alpha$ line (i.e.\ including the contribution from the Be disk) in logarithmic scale. The corresponding curve is shown by the red line for the Be disk with the default parameters. The magenta lines delimit the range of observed values of EW(Fe K$\alpha$) from 34 to 59\,eV. The cyan hatched area corresponds to the values of $N_{\rm H}$ towards the hot component as determined observationally \citep{Rau22}. These values range between $0.035\,10^{22}$\,cm$^{-2}$ and $1.38\,10^{22}$\,cm$^{-2}$. Figure\,\ref{coldens} shows that the EW model curve does not cross the intersection between the cyan and magenta areas.} Reproducing the observed EW range requires $n_{\rm H,A}$ of $4.7\,10^{14}$\,cm$^{-3}$ -- $10^{15}$\,cm$^{-3}$, whilst the highest observed absorbing column density restricts $n_{\rm H,A}$ to values below $8\,10^{13}$\,cm$^{-3}$. The highest value of the observed column density ($1.38\,10^{22}$\,cm$^{-2}$) was measured in an {\it XMM-Newton} observation in January 2021 when the X-ray spectrum of $\gamma$~Cas was extremely absorbed \citep{Rau22}. The second highest value ($1.16\,10^{22}$\,cm$^{-2}$) was observed in January 2022. However, those high values are thought to be due to transient clouds crossing our line of sight. Most of the time, the column density is at least a factor of 3 lower than in January 2021 \citep{Rau22}, which pushes the upper limit on $n_{\rm H,A}$ towards even lower values of about $2.5\,10^{13}$\,cm$^{-3}$. Therefore, in the propeller scenario with the default model parameters we obtain a mismatch by typically a factor of 20 between the values of $n_{\rm H,A}$ required to explain the EW(Fe K$\alpha$) and the values allowed by the observed column density\footnote{Assuming here that the entire observed column density stems from the cool shell.}.

A by-product of the above is that, as long as the reprocessing material remains optically thin, one would expect a linear relation between the column density of cool material towards the primary X-ray source and the EW of the fluorescent Fe K$\alpha$ line. Such a relation is indeed expected theoretically for a spherically symmetric distribution of the reprocessing material \citep[e.g.][]{Kal04}, and was empirically established for a sample of X-ray binaries \citep{Tor10}. Yet, no such correlation was found in multi-epoch X-ray data of $\gamma$~Cas \citep{Rau22}.

At this stage, it is important to point out that increasing the value $n_{\rm H,A}$ to match the strength of the observed fluorescent line would also impact the overall X-ray luminosity of the propelling NS. According to Eq.\ (6) of \citet{Pos17}, the X-ray luminosity of the hot shell scales with $n_{\rm H,A}^2$. Therefore, increasing $n_{\rm H,A}$ by a factor of $\sim 20$ would not only lead to a mismatch with the observed column density, but would also imply a theoretical X-ray luminosity that exceeds the observed one by a large factor.  

We tested whether extending the outer radius of the cool shell to the NS's Roche lobe radius would make a significant change. The answer is no: compared to the situation where the outer radius was set to $R_{\rm BHL}$, the EW of the line arising from the cool shell increases only by $ < 10$\%.

As we have shown in Sect.\,\ref{Bediskresult}, the contribution from the disk would be an order of magnitude higher if we adopt a flatter disk density profile \citep{Sil10} and assume the disk to extend out to the Be star's Roche lobe radius. {The dashed blue line in Fig.\,\ref{coldens} illustrates this situation. In this case, a range of $n_{\rm H,A}$ values between $2.7\,10^{14}$ and $7.0\,10^{14}$\,cm$^{-3}$ would allow a match between the total theoretical EW and the range of observed values.} Even for this extreme disk model, we still find a mismatch by an order of magnitude between the $n_{\rm H,A}$ densities required to explain the EW(Fe K$\alpha$) and the values of $n_{\rm H,A}$ allowed from the observed column density.

\section{Other $\gamma$~Cas stars \label{sect4}}
We now turn to the question whether other $\gamma$~Cas stars display Fe K$\alpha$ lines that agree with the predictions of the \citet{Pos17} model. For this purpose, we consider the cases of $\pi$~Aqr and $\zeta$~Tau. Though there are other $\gamma$~Cas stars that are brighter in X-rays (e.g.\ BZ~Cru), we focus here on objects with well-established orbital periods and Be disk density profiles. Indeed, a good knowledge of both properties is crucial for our modelling. Among the known $\gamma$~Cas stars that are established binary systems, $\pi$~Aqr has the shortest orbital period \citep[84.1\,d;][]{Bjo02}, thus illustrating a situation where the companion is closer to the Be disk, which should lead to an enhanced fluorescence by the disk material. On the other hand, $\zeta$~Tau illustrates the case of a shell star, where the system is seen in the plane of the Be disk. The parameters of these stars are discussed in Appendix\,\ref{paramother} and are summarized in Table\,\ref{tab_param}.
\subsection{$\pi$~Aqr}
Applying our code to the parameters of the $\pi$~Aqr system \citep[assuming an amplitude of 8.1\,km\,s$^{-1}$ for the Be star's RV curve,][]{Naz19a,Tsu23}, we obtained once again rather low EWs. The contribution from the Be disk would only amount to 0.32\,eV. Assuming $n_{\rm H,A} = 10^{13}$\,cm$^{-3}$, the cool shell would provide 0.75\,eV. The combined EW then amounts to 1.07\,eV, which is much smaller than the observed value of $68 \pm 8$\,eV \citep{Tsu23}, and such a weak line would simply remain undetected with current X-ray observatories. 

Adopting the same parameters, the value of the column density due to the cool shell would become $1.6\,10^{21}$\,cm$^{-2}$, which is close to the observed value ($2.6\,10^{21}$\,cm$^{-2}$). If the cool propeller shell were to account for most of the observed fluorescent line, and taking optical depth effects into account, the $n_{\rm H,A}$ density would have to be at least 120 times higher than our assumed value. This would then lead to a column density near $1.9\,10^{23}$\,cm$^{-2}$ towards the primary X-ray source. This value is 75 times higher than the observational column density, which de facto rules out the existence of such large values of $n_{\rm H,A}$. 

For the $\pi$~Aqr system, we note that $R_{\rm BHL} \simeq 38$\,R$_{\odot}$ is essentially identical to the radius of the NS's Roche lobe ($37.4$\,R$_{\odot}$), provided that the NS has a mass of 1\,M$_{\odot}$ (which is the case for the 8.1\,km\,s$^{-1}$ amplitude). If instead, we adopt the larger RV amplitude found by \citet{Bjo02} then the NS mass becomes 2.1\,M$_{\odot}$ and the Bondi-Hoyle-Lyttleton radius (80\,R$_{\odot}$) would exceed the NS's Roche lobe radius (47.3\,R$_{\odot}$) by a significant factor. Hence, if the larger RV amplitude holds, we would have to stop the integration at the NS's Roche lobe radius, and the EW contributed by the cool shell would thus only increase by 0.03\,eV reaching 0.35\,eV.

Likewise, if we allow the Be disk to extend out to the Roche lobe radius of the Be star, but keeping the same density profile as before ($n_0 = 0.45\,10^{13}$\,cm$^{-3}$, $\alpha = 2.5$), we obtain an EW of 0.45\,eV, which will not resolve our issue. Explaining a significant fraction of the observed fluorescent line strength by the effect of the disk, would typically require an increase in $n_0$ by at least an order of magnitude, as well as a flatter density profile. 

We thus conclude that, in the case of $\pi$~Aqr, we are left with the same dilemma as for $\gamma$~Cas: with the best estimates of the system characteristics it is simply not possible to explain the strength of the observed Fe K$\alpha$ line within the framework of the propelling NS scenario.

\subsection{$\zeta$~Tau}
Finally, we consider the $\zeta$~Tau system. Using our code with the default parameters yields an EW of the truncated Be disk contribution between 1.14 and 1.20\,eV depending on orbital phase. If instead the disk extended out to the Be star's Roche lobe, then it would contribute between 1.67 and 1.73\,eV. The mild phase dependence arises from the near edge-on inclination of the Be disk: depending on the orbital phase the Be star occults some parts of the disk illuminated by the primary X-ray source. Accordingly, the highest and lowest emission strengths would be observed respectively at inferior and superior conjunction of the NS. Likewise, adopting $n_{\rm H,A} = 10^{13}$\,cm$^{-3}$, the cool shell around the propeller would provide 0.70\,eV (or 0.73\,eV if the shell extends out to the NS's Roche lobe). This results in a combined EW of 1.84 -- 1.90\,eV if we consider the default parameters, which is again much smaller than the {observed strength of $\sim 55 \pm 8$\,eV (Naz\'e et al.\ in prep.)}.  

The associated column density towards the primary source due to the cool shell would be equal to $1.6\,10^{21}$\,cm$^{-2}$, which is much smaller than the value determined from the {\it Chandra} spectrum \citep[$1.55\,10^{23}$\,cm$^{-2}$,][]{Naz22} or from a recent {\it XMM-Newton} observation ($1.76\,10^{23}$\,cm$^{-2}$, Naz\'e et al.\ in prep.). Due to the orientation of the Be disk with respect to our line of sight, such an extremely high value could, to a large extent, reflect the absorption by the Be disk rather than the sole effect of the shell. In the context of our present study, where we wish to test the scenario of a primary X-source arising from the hot shell of a NS in the propeller stage, the observational column density could be reproduced by an increase in $n_{\rm H,A}$ by a factor of $\sim 100$ to $n_{\rm H,A} = 1.0\,10^{15}$\,cm$^{-3}$. Such a value of the density at the Alfv\'en radius would render the cool shell mildly optically thick. Accounting for the optical depth, we estimate an EW of 54.5\,eV for the corresponding cool shell contribution. Compared to an optically thin shell, the optical depth of the shell leads to a reduction of the EW by 22\%. {Whilst a value of $n_{\rm H,A} = 1.0\,10^{15}$\,cm$^{-3}$ could thus explain both the observed column density and EW of the Fe K$\alpha$ line, according to equation (6) of \citet{Pos17}, it would imply a drastic increase in the X-ray luminosity of the hot shell well beyond the observed value.}

\section{Discussion \label{sect5}}
X-ray spectra of $\gamma$~Cas stars display a prominent Fe line complex that includes a clear contribution of a fluorescent line due to low-ionization Fe atoms. Our calculations indicate that the model proposed by \citet{Pos17} struggles to account for the observed strengths of these fluorescent Fe K$\alpha$ lines. With the default parameters of the propeller model, the predicted line strength would be way too low for this line to be detectable in all three of the stars investigated in this study.

A way to overcome the problem would be to invoke geometrical effects that either prevent us from seeing the true level of the intrinsic primary X-ray emission or lead to the true column density of the reprocessing material being much higher than what is estimated from the observed photo-absorption column density. Whilst such a scenario could hold for a single system observed at a specific phase (e.g.\ similar to a high-mass X-ray binary in eclipse), it seems unlikely for three systems that have been observed at a variety of orbital phases and always yielded a similar global picture.

In our model, we consider reprocessing by material in the Be disk and in a cool shell around the propelling NS. One could speculate that the polar wind of the Be star plays some role in the fluorescent process. Using International Ultraviolet Explorer (IUE) spectra, \citet{Pri89} detected a wind signature in the C\,{\sc iv} $\lambda\lambda$\,1548, 1550 doublet of $\gamma$~Cas. From a fit to the line profile, he inferred a terminal wind velocity of 1500\,km\,s$^{-1}$ and a value of $\log{[\dot{M}\,q({\rm C}^{3+})]} = -10.46$ (in M$_{\odot}$\,yr$^{-1}$). Assuming the same ionization fraction in Be-star winds as in O-star winds, this value would correspond to a mass-loss rate of $3.8\,10^{-8}$\,M$_{\odot}$\,yr$^{-1}$. \citet{Smi99} and \citet{Cra00} measured a higher terminal velocity of 1800\,km\,s$^{-1}$ from the blue edge of discrete absorption components in the Si\,{\sc iv} $\lambda\lambda$\,1394, 1403 lines observed in \textit{Hubble} Space Telescope spectra. Adopting a mass-loss rate of $5\,10^{-8}$\,M$_{\odot}$\,yr$^{-1}$ and a wind velocity of 1500\,km\,s$^{-1}$, a generous upper limit on the electron density at the position of the NS is $\sim 1.7\,10^6$\,cm$^{-3}$ \citep[see also][]{Cra00}. This is about 50 times lower than the density of the cool shell at its outer boundary, $R_{\rm BHL}$, assuming $n_{\rm H,A} =  10^{13}$\,cm$^{-3}$. Hence, this polar wind is extremely unlikely to play any significant role in the formation of the fluorescent line.

The results of our calculations depend on the atomic parameters that were used. Atomic parameters used in astrophysics are subject to uncertainties, and the quantities used in our calculations are no exception to that rule. Comparing Fe K fluorescent transition probabilities computed with different approximations, \citet{Pal03} concluded that the atomic parameters are likely accurate to 10\%. To increase the strength of the synthetic lines up to the level of the observed lines, one would need to increase the fluorescence yields and cross-section by much more than this 10\%. Moreover, fluorescence yields are by definition lower than 1, but to match the observed line strengths, one would have to increase them by more than a factor of 3, which would make them exceed this upper limit.

Finally, one could ask whether the usage of solar Fe abundance \citep{Asp09} is appropriate. A higher than solar Fe abundance could indeed alleviate the mismatch between calculations and observations. Yet, we need to keep in mind that fluorescent Fe K$\alpha$ is not the only Fe line in the X-ray spectrum of $\gamma$~Cas. The primary X-ray spectrum also displays several Fe\,{\sc xvii} lines around 0.7 -- 0.8\,keV as well as the Fe\,{\sc xxv} triplet near 6.7\,keV and the Fe\,{\sc xxvi} Ly$\alpha$ lines at 6.97\,keV. These features help constrain the Fe abundance in global fits of the X-ray spectrum. Such fits do not provide any evidence of higher than solar abundance, and instead yield a sub-solar Fe abundance \citep{Smi04,Lop10,Rau22}, which would only increase the discrepancy. 

\citet{Tor10} analysed Fe K fluorescent lines in {\it Chandra} High Energy Transmission Grating high-resolution spectra of a sample of X-ray binaries, including nine high-mass X-ray binaries (HMXBs) and $\gamma$~Cas. They established that the wavelengths of the fluorescent Fe K$\alpha$ lines indicate that they arise from Fe ions up to Fe\,{\sc x}. \citet{Tor10} further found that the lines are narrow (FWHM $< 5$\,m\AA), which indicates that the reprocessing material is unlikely to rotate at high velocities (as would be the case for the inner regions of an accretion disk close to an accreting NS). The empirical relation between the EW(Fe K$\alpha$) and the column density of the absorbing material suggests a roughly spherical distribution of the reprocessing material. \citet{Tor10} also showed that X-ray binaries display a so-called X-ray Baldwin effect, that is to say, an anti-correlation between the X-ray luminosity and EW(Fe K$\alpha$). This result was confirmed by \citet{Gim15} based on an independent sample of the {\it XMM-Newton} European Photon Imaging Camera (EPIC-pn) spectra of HMXBs. This effect was interpreted as a decrease in the volume of reprocessing material due to increasing ionization when the X-ray luminosity increases. In the propeller scenario for $\gamma$~Cas stars, one could expect a very similar behaviour, as the boundary between the hot and cool shell would move farther out in response to an increase in the X-ray luminosity, thereby decreasing the available volume of reprocessing material. Quite interestingly though, this effect is not observed in $\gamma$~Cas stars as a group \citep{Gim15} nor in $\gamma$~Cas itself when considering observations from different epochs \citep{Rau22}. This suggests that the conditions for fluorescence are fundamentally different in supergiant HMXBs and $\gamma$~Cas stars.     

We thus conclude that the observed strength of Fe K$\alpha$ in $\gamma$~Cas stars is at odds with the propeller NS scenario. This is actually not the only problem that this scenario encounters. \citet{Smi17} previously refuted the propeller scenario on other grounds. Indeed, it does not account for the observed correlations between the UV and X-ray emission of $\gamma$~Cas  because the density assumed by \citet{Pos17} at the Alfv\'en radius is too low compared to the density required to explain the very short decay time of the shots seen in the X-ray light curve. Indeed, the X-ray emission of $\gamma$~Cas stars varies over many different timescales, from a few seconds to months and years. The most rapid variations take the form of shots on timescales as short as 4\,s \citep{Smi98}. Such short decay times require densities of $> 10^{14}$\,cm$^{-3}$, which exceed those assumed by \citet{Pos17} for the shell of the propeller NS. Finally, based on evolutionary considerations, \citet{Smi17} show that the short duration of a propeller phase is unlikely to account for the number of $\gamma$~Cas stars compared to the population of known Be HMXBs. This argument has been reinforced over the last few years as the number of known $\gamma$~Cas stars has considerably increased since 2017 \citep{Naz18,Naz20,Naz22,Naz23}. Therefore, in view of the various results and considerations in this paper as well as the other problems identified by \citet{Smi17}, we can now answer the question raised in the title of this article: it seems very unlikely that the $\gamma$~Cas phenomenon traces a population of Be stars orbited by a propelling NS.  

\begin{acknowledgements}
  GR thanks Dr.\ Patrick Palmeri for discussion about the details of the fluorescence mechanism, as well as Jahanvi and Drs.\ Ya\"el Naz\'e, Myron Smith and Christian Motch for numerous discussions about various aspects of $\gamma$~Cas stars. This research is supported by the BELgian federal Science Policy Office (BELSPO) and the European Space Agency through the PRODEX HERMeS grant. 
  \end{acknowledgements}

\appendix
\section{Stellar and disk parameters of $\pi$~Aqr and $\zeta$~Tau \label{paramother}}
In this appendix we provide an overview of the literature on $\pi$~Aqr and $\zeta$~Tau, which is relevant for the input parameters needed for our model (see Table\,\ref{tab_param}).

Pi Aqr (also known as HD~212571) is classified as a B1\,III-IVe star. The $\gamma$~Cas nature of its X-ray emission was discovered by \citet{Naz17}. The hot plasma component has a temperature of $11.5$\,keV \citep{Naz17,Naz18}, and the flux (corrected for interstellar absorption) amounts to $1.07\,10^{-11}$\,erg\,cm$^{-2}$\,s$^{-1}$ in the 0.5 -- 10\,keV energy range. The circumstellar hydrogen column density was found to be $0.26\,10^{22}$\,cm$^{-2}$ \citep{Naz17}. The EW of the Fe K$\alpha$ line is reported at a value of $(68 \pm 8)$\,eV by \citet{Tsu23}. \citet{Bjo02} observed $\pi$~Aqr when the disk had cleared away and showed that it is a binary with an essentially circular orbit with a period of 84.1\,d. These authors obtained an SB2 orbital solution with a semi-amplitude of 16.7\,km\,s$^{-1}$ for the Be star, and 101.4\,km\,s$^{-1}$ for the weak H$\alpha$ emission line that moved in anti-phase with the absorption line of the B1 star. Their results imply dynamical masses of $m_{\rm Be}\,\sin^3{i} = 12.4$\,M$_{\odot}$ and $m_{\rm comp}\,\sin^3{i} = 2.0$\,M$_{\odot}$ \citep{Bjo02}. However, more recent studies, performed when the disk had re-developed, led to a significantly lower semi-amplitude of the Be star's orbital motion of about 8.1\,km\,s$^{-1}$ \citep{Naz19a,Naz19b,Tsu23}. This situation implies considerable uncertainties on the masses of the components: the Be star's dynamical mass $m_{\rm Be}\,\sin^3{i} = 12.4$\,M$_{\odot}$ obtained by \citet{Bjo02} was converted into a mass of $(14 \pm 1)$\,M$_{\odot}$ by \citet{Zha13}, whilst \citet{Cat13} and \citet{Naz19b} respectively estimate 13 and $(11 \pm 1.5)$\,M$_{\odot}$. Adopting the lower semi-amplitude of the Be star's RV curve, the dynamical masses become $m_{\rm Be}\,\sin^3{i} = 10.6$\,M$_{\odot}$ and $m_{\rm comp}\,\sin^3{i} = 0.85$\,M$_{\odot}$. We adopt here a mass of the Be star of 12.7\,M$_{\odot}$, which is the mean of the above estimates. Kepler's third law then implies a semi-major axis of 193\,R$_{\odot}$ for the lower RV semi-amplitude and 198\,R$_{\odot}$ for the higher RV semi-amplitude. For the stellar radius, \citet{Zha13} estimated 13.0\,R$_{\odot}$ whilst other literature estimates rather fall in the range 5.9 to 7.7\,R$_{\odot}$ \citep{Cat13,Arc18,Naz19b}. The larger radius proposed by \citet{Zha13} most likely stems from their higher distance estimate of 740\,pc, whilst {\it Hipparcos} and {\it Gaia} parallaxes rather suggest a distance around 240\,pc \citep{Naz18,Arc18}. Here, we adopt a value of 6.5\,R$_{\odot}$ for the stellar radius. This then leads to a Keplerian velocity of 610\,km\,s$^{-1}$ at the inner edge of the Be disk. Effective temperatures between 23.3 and 27.6\,kK are quoted in the literature \citep{Bjo02,Fre05,Tou10,Zha13,Arc18,Naz18,Naz19b}. We take the mean of 25\,kK. Orbital and/or disk inclinations are estimated between 50 -- 75$^{\circ}$ \citep{Bjo02}, $45^{\circ}$ \citep{Sil10}, 65 -- 85$^{\circ}$ \citep{Zha13}, 60$^{\circ}$ \citep{Arc17}, 50 -- 75$^{\circ}$ \citep{Naz19b}, and $(56 \pm 5)^{\circ}$ \citep{Sig23}. The lowest values would lead to an unusually large absolute dynamical mass for the Be star. We thus adopt a value of 70$^{\circ}$ consistent with the estimated dynamical minimum mass and our above mass estimate for the B1 star. The density at the inner edge of the disk was evaluated to be $10^{-11}$\,g\,cm$^{-3}$ \citep{Sil10}, $6\,10^{-13}$\,g\,cm$^{-3}$ \citep{Cat13},  $7.5\,10^{-12}$ -- $10^{-11}$\,g\,cm$^{-3}$ \citep{Arc17}. This rather wide range reflects at least partially the epoch-dependence of the disk strength which has increased significantly over the last decade \citep{Naz19a,Tsu23}. The values of \citet{Arc17} were obtained at epochs close to the {\it XMM-Newton} observation. Hence, we adopt here a value of $10^{-11}$\,g\,cm$^{-3}$ which translates into an hydrogen number density of $4.5\,10^{12}$\,cm$^{-3}$. The disk density power-law exponent $\alpha$ of $\pi$~Aqr ranges between 2.5 and  3.5 \citep{Sil10,Cat13,Arc17}. We adopted $\alpha = 2.5$, the value from \citet{Arc17}. With $q = 0.165$ (higher RV amplitude) or $q = 0.080$ (lower RV amplitude), the Roche lobe radius of the Be star is estimated respectively to 106.3 or 115.0\,R$_{\odot}$. Likewise, for the companion we obtain a Roche lobe radius of 47.3 or 37.4\,R$_{\odot}$. Finally, the 3:1 resonance truncation radius of the Be disk amounts to 91\,R$_{\odot}$.\\

Zeta Tau (= HD~37202) is a B1\,IVe shell star. A first estimate of the characteristics of its X-ray spectrum was derived by \citet{Naz22} from a heavily piled-up {\it Chandra} spectrum. A plasma temperature of $11.1 \pm 4.3$\,keV was inferred, along with a flux of $1.66\,10^{-11}$\,erg\,cm$^{-2}$\,s$^{-1}$ evaluated over the 0.5 -- 10\,keV band and corrected for ISM absorption. The circumstellar absorption column was found to be rather high with $(15.5 \pm 2.0)\,10^{22}$\,cm$^{-2}$. A recent, pile-up free, {\it XMM-Newton} observation yielded a plasma temperature of $9.0 \pm 0.3$\,keV, an ISM-corrected flux of $1.76\,10^{-11}$\,erg\,cm$^{-2}$\,s$^{-1}$ and a circumstellar column density of $(17.6 \pm 0.2)\,10^{22}$\,cm$^{-2}$ (Naz\'e et al.\ in prep.). The mass of the Be star is estimated close to 11.2\,M$_{\odot}$ \citep{Gie07,Car09,Ruz09,Tou11}. For the mean stellar radius, values of 5.5\,R$_{\odot}$ \citep{Gie07,Tou11} or 6.1\,R$_{\odot}$ \citep{Coc19} are given in the literature. \citet{Car09} accounted for the rotational flattening and rather proposed 5.9 and 7.7\,R$_{\odot}$, respectively, for the polar and equatorial radii. Since we use the stellar radius as the inner radius of the disk, which is assumed to lie in the equatorial plane, we adopt a radius of 7.7\,R$_{\odot}$. These numbers lead to an orbital velocity of 525\,km\,s$^{-1}$ at the inner edge of the circumstellar disk. Mean effective temperatures were derived in the range from 19.0 to 20.1\,kK \citep{Gru06,Gie07,Coc19}. Accounting for gravity darkening, \citet{Car09} inferred 18.0 and 25.0\,kK, respectively, at the equator and at the pole. We adopt here a value of 20\,kK. The inclination of the Be disk is usually given in the range from $75^{\circ}$ to $89^{\circ}$ \citep{Gru06,Gie07,Car09,Tou11} with the exception of \citet{Coc19} who rather quote an inclination of $67^{\circ}$. We adopt a value of $85^{\circ}$, consistent with the shell aspect of the Balmer lines. Estimates of the density at the inner edge of the disk range from $5.9\,10^{-11}$\,g\,cm$^{-3}$ to $1.9\,10^{-10}$\,g\,cm$^{-3}$ \citep{Gie07,Car09,Tou11}. The mean value $1.3\,10^{-10}$\,g\,cm$^{-3}$ translates into a hydrogen number density of $5.8\,10^{13}$\,cm$^{-3}$. The $\alpha$ exponent ranges between 2.9 and 3.5 \citep{Gie07,Car09,Tou11,Kle19} and we adopt a value of 3.0. The $p$ exponent was assumed to be 0 by \citet{Car09}. $\zeta$~Tau has an orbital period of 132.987\,d and a likely companion mass of 0.87 -- 1.02\,M$_{\odot}$ \citep{Ruz09}, which results in a mass ratio $q = 0.086$. Kepler's third law then yields an orbital separation of 252\,R$_{\odot}$, and the Roche lobe radii are estimated to 148.7\,R$_{\odot}$ for the Be star and 49.9\,R$_{\odot}$ for its companion. The 3:1 truncation radius of the Be disk amounts to 118\,R$_{\odot}$. The semi-amplitude of the Be star's orbital motion is 7.4\,km\,s$^{-1}$ \citep{Ruz09,Naz22}. For the distance of $\zeta$~Tau, we adopt 136\,pc \citep{Naz22}.
\end{document}